\documentclass[]{mnras} 
\usepackage{graphicx}	
\usepackage{bm}  
\usepackage{color}
\usepackage[T1]{fontenc} 
\DeclareRobustCommand{\VAN}[3]{#2}
\let\VANthebibliography\thebibliography
\def\thebibliography{\DeclareRobustCommand{\VAN}[3]{##3}\VANthebibliography}
\usepackage{graphicx}	
\usepackage{amsmath}	
\usepackage{amssymb}	

\def\be{\begin{equation}} \def\ee{\end{equation}}
\def\Ha{H$\alpha$} \def\ha{H$\alpha$} 
\def\Tb{T_{\rm b}} \def\Te{T_{\rm e}}
\def\deg{^\circ} 

\def\Iha{I_{\rm H\alpha}}
\def\Ira{I_{\rm 1.4\ GHz}} 
\def\Iunit{erg cm$^{-2}$ s$^{-1}$ sr$^{-1}$}  
\def\Sigunit{erg cm$^{-2}$ s$^{-1}$ sr$^{-1}$ Hz$^{-1}$}
\def\EMunit{cm$^{-6}$ pc } 
\def\EM{\red{\rm EM}} 
\def\IhaNPS{\Iha ({\rm NPS})}
\def\Q{\mathcal{Q}} 

\def\red{}

\begin{document}   

\title[\Ha\ faintness of the North Polar Spur]
{On the \Ha\ faintness of the North Polar Spur}
\author[Y. Sofue et al.]{Yoshiaki {\sc Sofue}$^{1}$\thanks{E-mail: sofue@ioa.s.u-tokyo.ac.jp}, Jun {\sc Kataoka}$^2$ and Ryoji {\sc Iwashita}$^2$
\\
1. Institute of Astronomy, The University of Tokyo, Mitaka, Tokyo 181-0015, Japan \\
2. Faculty of Science and Engineering, Waseda University, Shinjyuku, Tokyo, 169-8555, Japan
} 
\maketitle
\begin{abstract}   
The ratio of \Ha\ intensity to 1.4 GHz radio continuum intensity in the North Polar Spur (NPS) is measured to be $\lesssim 50$, two orders of magnitude smaller than the values of $\sim 10^4$ observed in the typical shell-type old supernova remnants, Cygnus Loop and S147.
The extremely low \Ha-to-radio intensity ratio favours the Galactic-Centre explosion model for NPS, which postulates a giant shock wave at a distance of several kilo parsecs in the hot and low-density Galactic halo with low hydrogen recombination rate, over the local supernova(e) remnant model.  
\end{abstract}
 

\section{Introduction} 

The North Polar Spur (NPS) forms the northeastern edge of the giant Galactic bubble, composing the Loop I of radio continuum \cite{haslam+1982} and X-ray emissions \cite{snowden+1997,predehl+2020} with the brightest ridge at $(l,b)\sim(30\deg,20\deg)$ \cite{sofue+1979}.
Due to the sharp-edged shell-like morphology, the NPS is interpreted as a spherical shock wave driven by an explosive energy release at the loop centre.

There are two ideas to explain the origin of the explosion.
One is the explosion(s) of nearby supernova(e)
\cite{hanbury+1960,berk+1971,egger+1995,aschenbach+1999,wolleben2007,dickinson2018}
in the Sco-Cen OB Associations at a distance of $\sim 140$ pc \cite{dezeeuw+1999}.
This hypothesis assumes supernova remnant(s) expanding in the low temperature, high density ISM at $kT\sim 0.01-1$ eV ($T\sim 10^2-10^ 4$ K), $n\sim 1$ H cm$^{-3}$, and height $ |z|\lesssim 30$ pc inside the Galactic disc.
However, when most of nearby shell-type SNRs with angular diameters $\gtrsim 1\deg$ were optically identified by red (\Ha)-sensitive emulsions \cite{vandenbergh+1973}, the absence of \Ha\ counterpart to NPS, if it is the closest SNR(s), has been a mystery for over half a century \cite{sofue+1974}.
In fact, this problem was already {pointed out} in the earliest paper which suggested the SNR idea for the first time \cite{hanbury+1960}.

The other idea is {an explosion in the Galactic nucleus} or a starburst in the Galactic Centre (GC) \cite{sofue1977,sofue+2016,kataoka+2018}, which postulates a shock wave propagating in the hot, low-density halo with $kT\sim 0.2$ keV ($\sim 10^{6.3}$ K) and $n\sim 10^{-3}$ H cm$^{-3}$ at $z\sim 3-8$ kpc.
Because of the high temperature the gas is almost perfectly ionized and the hydrogen recombination is limited, so that the \ha\ absence may not contradict this idea.

In the present paper, we revisit this classical issue of NPS's optical dimness, which appears to have not been explored by quantitative analysis based on the observational data.
We examine the ratio of \Ha\ intensity to radio continuum intensity in the NPS and in the most typical shell-type SNRs, Cygnus Loop and S147 \cite{vandenbergh+1973}, and 
focus on the difference in the radiation processes in the shock fronts expanding into the Galactic halo and into the disc.
 
\begin{figure*} 
\begin{center}      
\includegraphics[width=15cm]{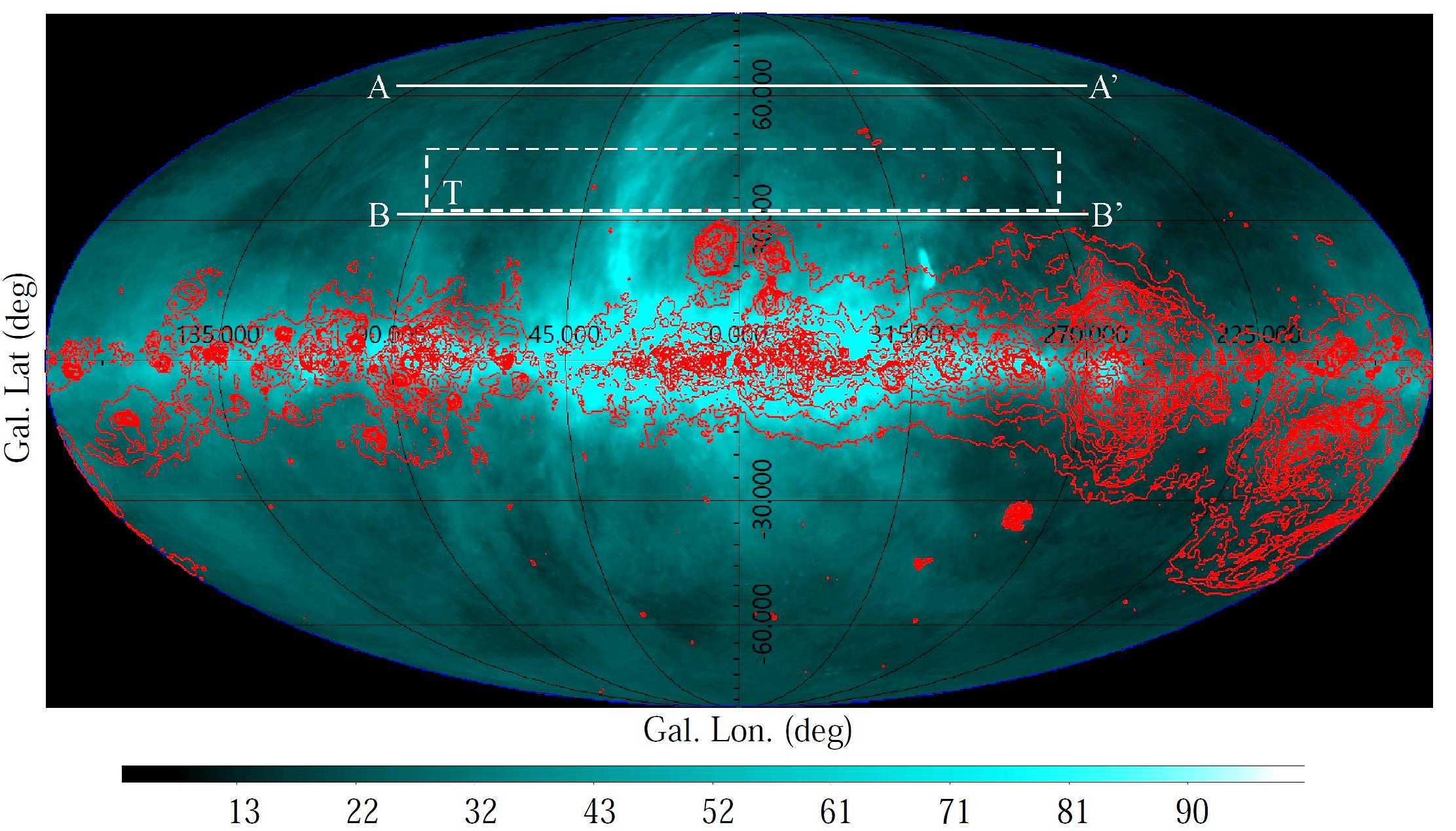}  
\hskip -2cm $\Tb$ (K) \\
\end{center} 
\caption{All-sky map of the 408 MHz brightness temperature ($\Tb$) from the Bonn-Parkes survey (Haslam et al. 1982)  
overlaid by contours of the \Ha\ intensity (at 5, 12, 19, 27, 39, 56R, ...) from the WHAM survey (Haffner et al. 2003, 2010).
Inserted box and lines are used for analyses in Fig. \ref{fig-cross} and \ref{fig-tt}. 
{Angular resolution of the radio map is $0\deg.85$ and \ha\ contours are smoothed to $0\deg.45$. Background filtering is not applied here.} }
\label{fig-allsky}    
\end{figure*}

\section{Optical vs radio continuum emissions}

\subsection{Data and measured quantities}
 \label{data}
 
The radio continuum data were taken from the 408 MHz Bonn-Parkes 
\cite{haslam+1982}, 1420 MHz Stockert-Villa Elisa 
\cite{reich+2001} all-sky survey, and 1420 MHz Bonn 100-m Galactic plane survey 
\cite{reich+1990,reich+1997}.
Also, radio surveys at 
22 \cite{22mhz},
150 \cite{150mhz}, and
820 \cite{820mhz}.
We use the brightness temperature $\Tb$ in K which is related to the surface brightness by $\Sigma=2k \Tb/\lambda^2$ in \Sigunit.
We also use the integrated intensity at $\nu=1.4$ GHz defined by
\begin{equation}
    I_\nu=\nu \Sigma
\end{equation} 
in unit of \Iunit\ in order to compare with the \ha\ intensity.
The \Ha\ data were taken from the Wisconsin \Ha\ all-sky map (WHAM) 
\cite{haffner+2003,haffner+2010}, and the intensity unit is Rayleigh (R) defined by
\be 
{\rm 1R}= 
\frac{10^6}{4\pi} \frac{\rm photons}{\rm cm^{2} \ s\ sr}  
=2.4\times 10^{-7} \frac{\rm erg}{\rm cm^{2} \ s \
sr}
\ee
at \Ha, and 1 R corresponds to emission measure of
$\EM= 2.25$ \EMunit for gas at a temperature of
8000 K \cite{haffner+2003}.

As we are interested in the \Ha-to-radio continuum intensity ratio, we introduce a parameter, $\Q$, defined by
\begin{equation}
    \Q=\Iha/\Ira. \label{eqHRatio}
\end{equation} 
Here, the radio brightness at $\Tb=1$ K corresponds to $\Ira=8.76\times 10^{-10}$ \Iunit.
So, a region with $\Iha=1$ R and $\Tb=1$ K has the intensity ratio of $\Q=274$.

\subsection{All-sky maps}

Figure \ref{fig-allsky} shows the all-sky radio brightness ($\Tb$) map at 408 MHz in the $(l,b)$ coordinates obtained from the Bonn-Parkes survey \cite{haslam+1982} overlaid with contours of the \Ha\ intensity ($\Iha$) map from the WHAM survey \cite{haffner+2003}.
The FWHM resolution of the 408 MHz map is $0\deg.85$.
The \ha\ data had a resolution of $6'$, while the contours here are drawn after smoothing to the same resolution as radio.  
We immediately notice that Loop I, including NPS, is invisible or very weak in \ha\ despite the high radio brightness even toward the most prominent ridge at $l\sim 30\deg$ and $b\sim 20 - 70\deg$.
 
\subsection{{Perpendicular optical filaments in the Aquila Rift}}

\begin{figure} 
\begin{center}   
\includegraphics[width=8cm]{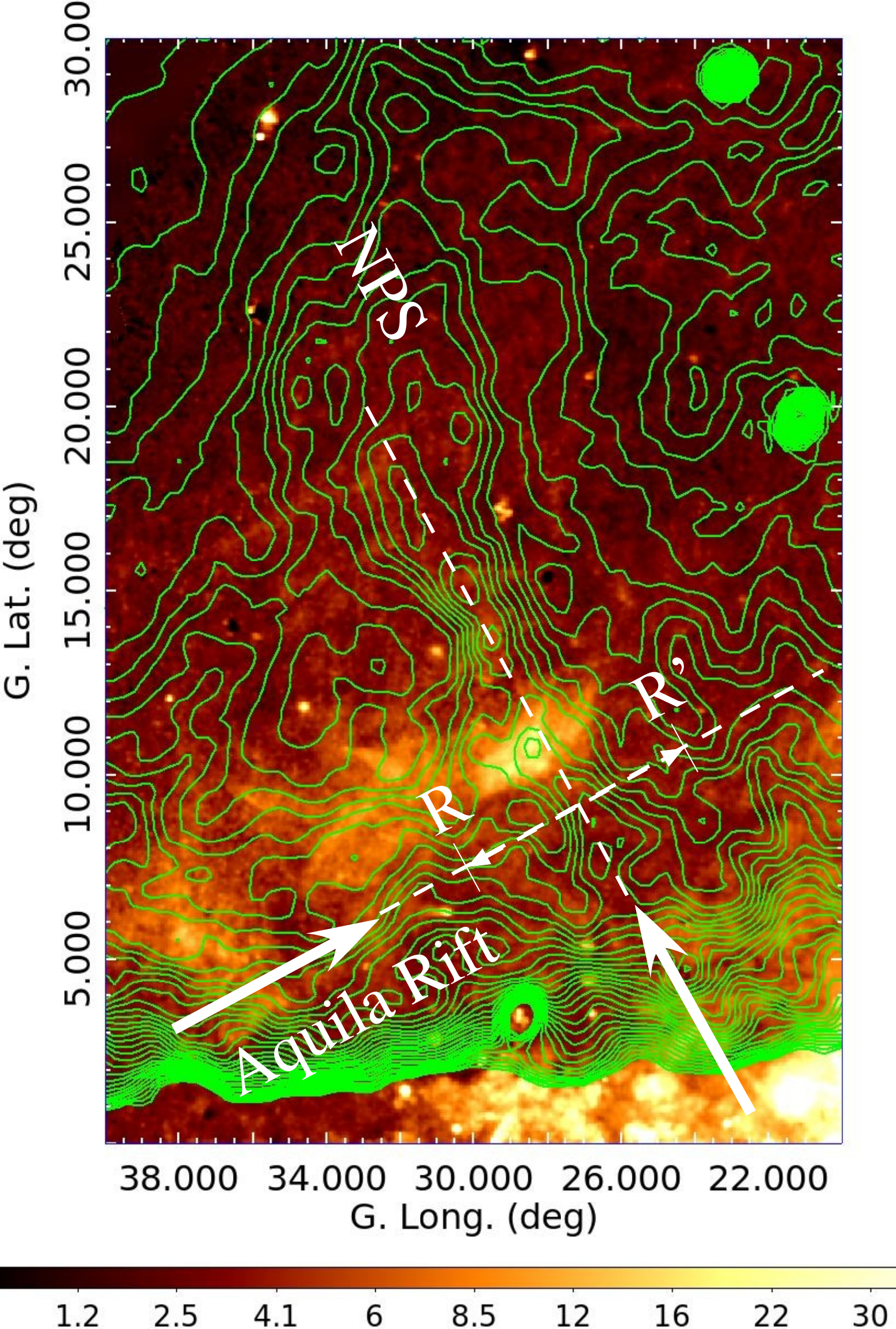} \\ 
\includegraphics[width=8cm]{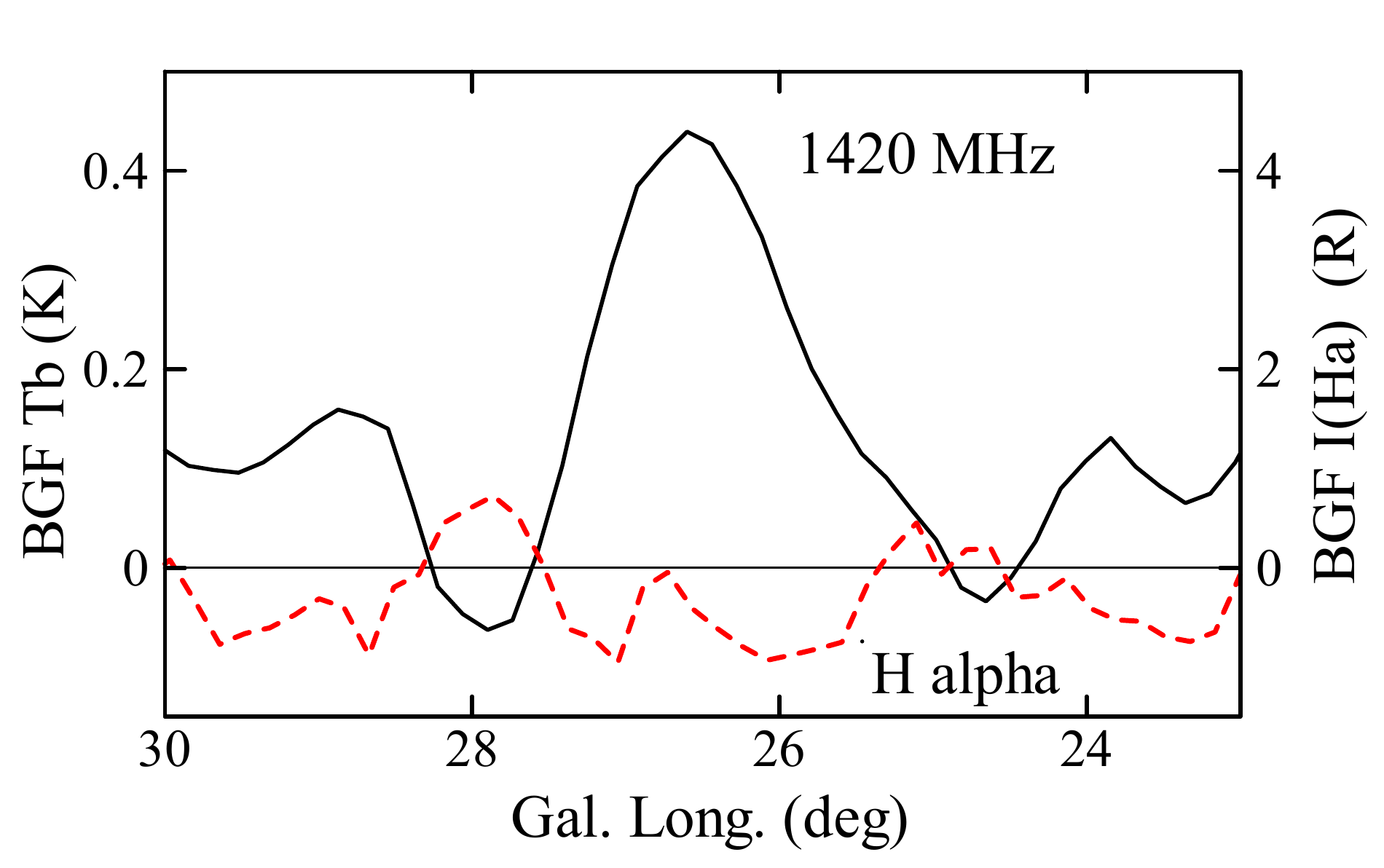}  
\end{center}  
\caption{[Top] The brightest ridge of NPS at 1420 MHz by contours (interval 0.1 K; maximum contour $\Tb=8$ K) from the Stockert survey (Reich et al. 2001) overlaid on the \ha\ map (scale is non linear by the bar). 
Many \ha\ filaments run parallel to the Aquila Rift (PA$\sim 120\deg$), but they are perpendicular to the NPS (PA$\sim 30\deg$). No {background subtraction} is applied in this figure.
[Bottom] Cross section of 21-cm and \ha\ along line R-R'. 
Here, the smooth emission has been removed by the background-filtering (BGF) technique (Sofue and Reich 1979).
(See section \ref{sec-spectra} for the description of BGF.)}

\label{fig-nps-up}  
\end{figure}

Figure \ref{fig-nps-up} enlarges the radio map at 1420 MHz \cite{reich+2001} around the brightest region of the NPS by contours superposed on the \Ha\ intensity map.  
The figure shows several \ha\ filaments running parallel to the Aquila Rift, which is the dark absorption belt from $(35\deg,0\deg)$ to $(20\deg,13\deg)$, with $\Iha\sim 2-5$ R at position angle of PA$\sim 120\deg$, but they cross the NPS at right angle.
The perpendicular orientation indicates that these \ha\ filaments are not related to NPS. 
We also stress that the darkest part of the Rift at $(l,b)\sim (27\deg,8\deg)$ shows no enhancement across NPS as shown by a comparison of the cross sections of radio and \ha\ intensities in the lower panel of Fig. \ref{fig-nps-up}.
{This means that no \ha\ feature associated with NPS exists in front of the Aquila Rift whose distance is $\sim 430$ pc \cite{sofue2015,sofue+2017}, which contradicts the local origin model for Loop I.
However, it does not necessarily deny the possibility that there is \ha\ associated with NPS, if it is located behind the Rift. } 
In either case, 
it is difficult to determine how much of the \ha\ feature is associated with the NPS.
We therefore exclude this region from quantitative measurements, and restrict our analysis to the spur at higher latitudes than $b\sim 30\deg$.

\subsection{Intensity profiles}

\begin{figure}  
\begin{center} 
A - A' \\
\includegraphics[width=8cm]{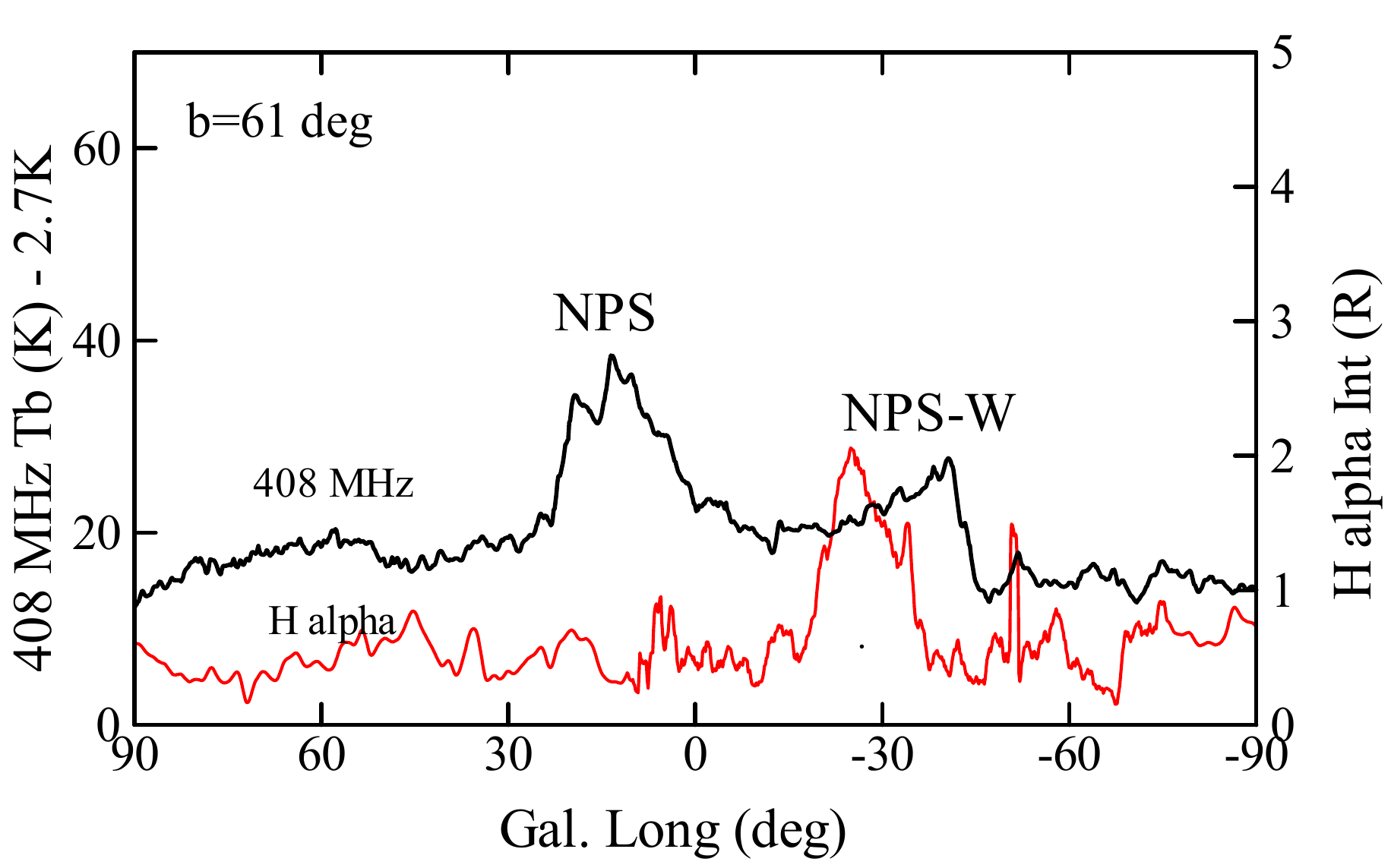}  \\
B - B' \\
\includegraphics[width=8cm]{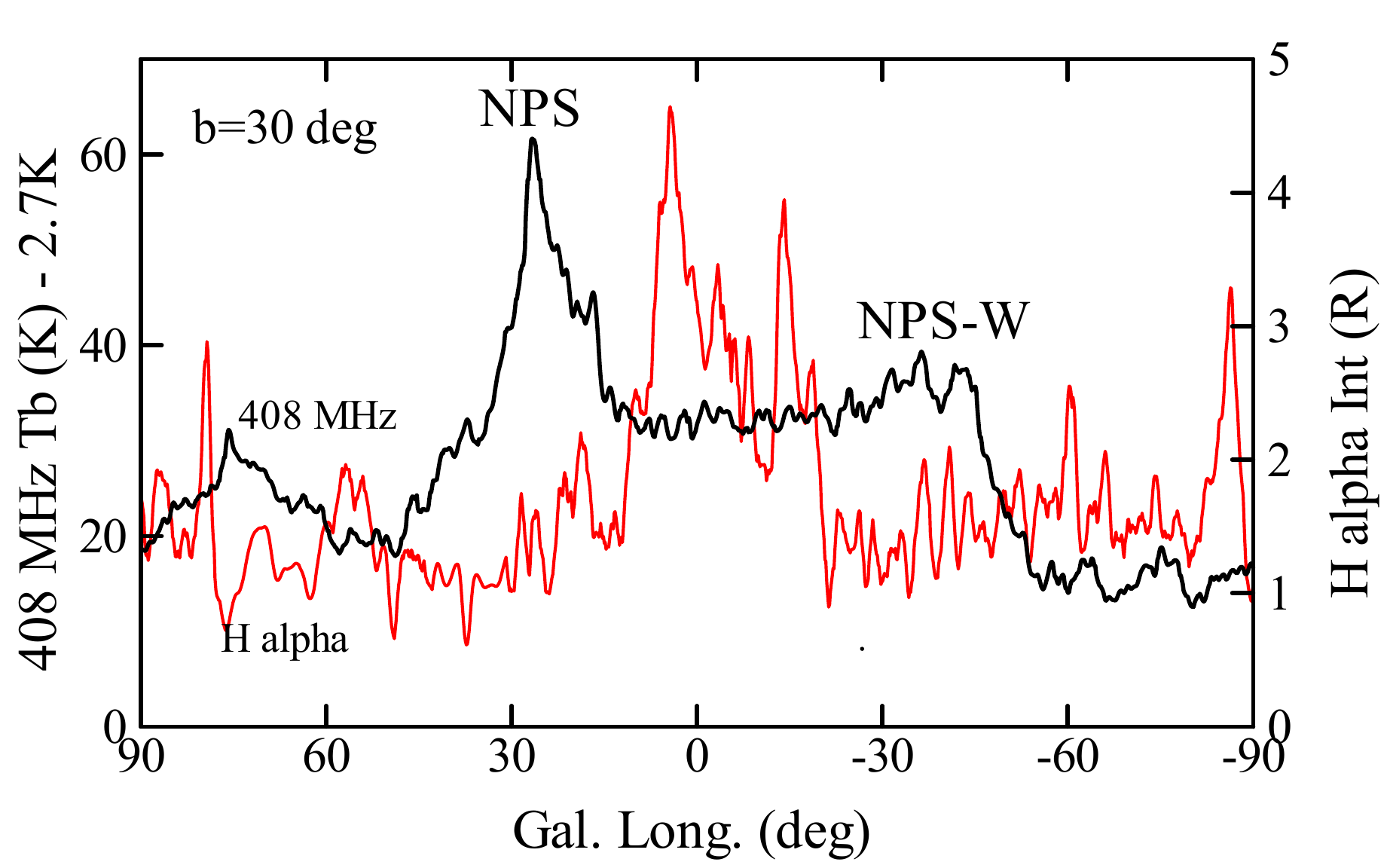} 
\end{center} 
\caption{Horizontal cross sections of the 408 MHz radio continuum (black line) and \Ha\ (thin red line) along the lines A-A' and B-B' in Fig. \ref{fig-allsky} at $b=30\deg$ and $61\deg$. Note the lack of \Ha\ emission corresponding to the radio peaks of NPS and NPS-west.
{Note that \ha\ enhancements around $l\sim -30\deg$ along A-A' and $l\sim 0\deg$ along B-B' are foreground or background} diffuse HII regions not related to the NPS.
}
\label{fig-cross}  
\end{figure}

Fig. \ref{fig-cross} shows horizontal cross sections of the 408 MHz radio brightness and \Ha\ intensity across the NPS and NPS West, or across the Loop I, at $b=61\deg$ and $30\deg$ along lines AA' and BB' in Fig. \ref{fig-allsky}, respectively.
The radio profiles exhibit a typical double-horn structure indicative of the intensity variation across a spherical shocked shell.
On the other hand, the \Ha\ intensity distribution is almost flat except for the broad enhancements by local HII regions ($l\sim -25\deg$ at $b=61\deg$ and $l\sim 0\deg$ at $b=30\deg$), which are unrelated to the NPS.

\subsection{Intensity-intensity plots}

\begin{figure}  
\begin{center} 
\includegraphics[width=8cm]{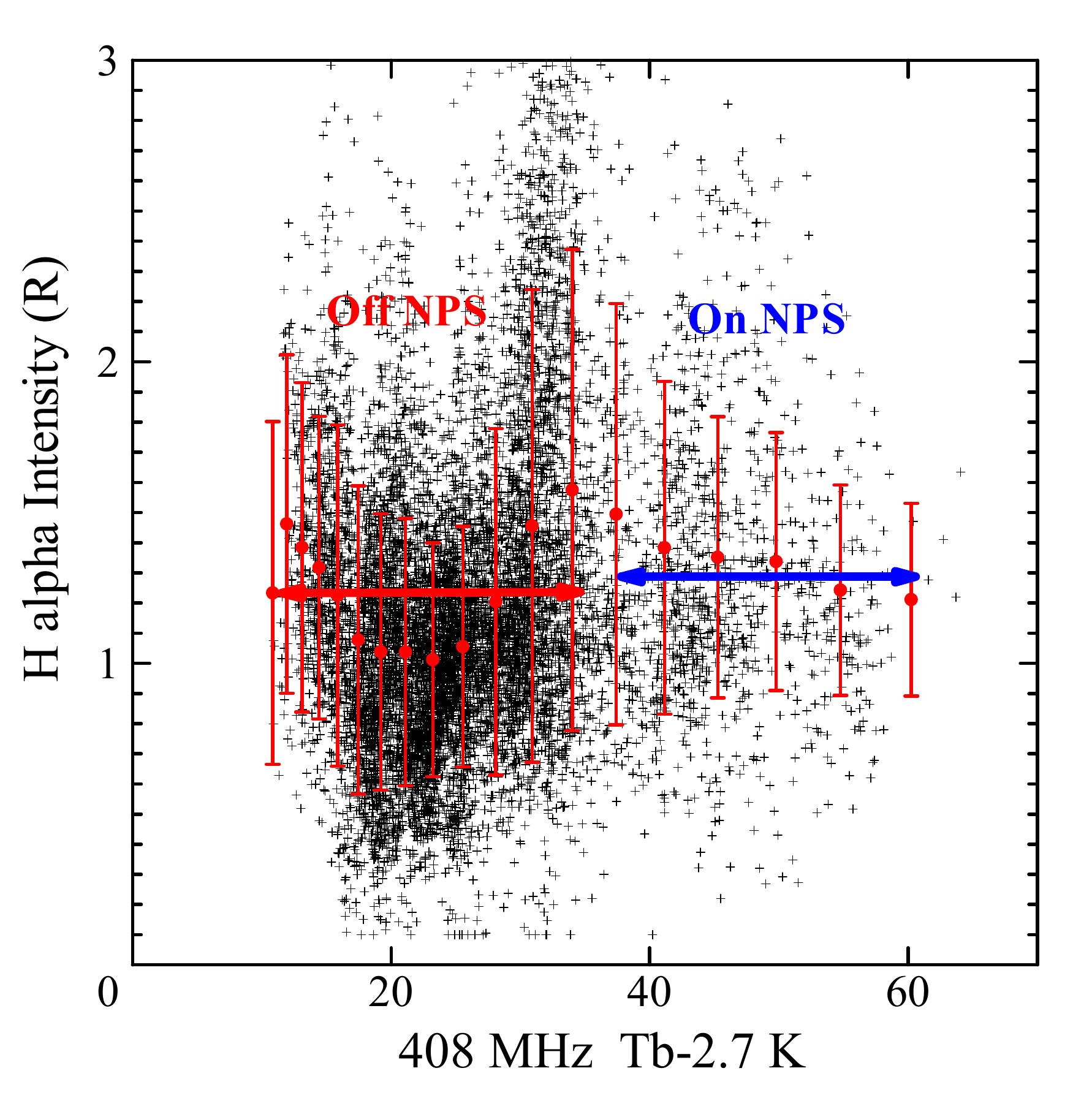}    
\end{center} 
\caption{$TT$ Plot of the \Ha\ intensity against 408 MHz brightness temperature $\Tb$ after subtracting 2.73 K in the area 'T' of Fig. \ref{fig-allsky} (from $l\sim270\deg$ to $90\deg$ and $b=+30\deg$ to $+45\deg$). 
The blue and red arrows indicate the on-NPS and off-NPS emissions, respectively.
{The background-filtering is not applied here.
} 
}
\label{fig-tt}  
\end{figure}

The faintness of \Ha\ emission in NPS can be more quantitatively displayed by plotting the \Ha\ intensity against radio intensity in and around the objects using the so-called $TT$ plots.
Fig. \ref{fig-tt} shows \Ha\ intensity in the squared area 'T' of Fig. \ref{fig-allsky} across the NPS-E (east) and NPS-W (west) plotted against 408 MHz $\Tb$ after subtracting the 2.7 K cosmic background emission.
Open circles are running means around individual centres of the $\Tb$ bins with the bars indicating the dispersion in \Ha\ intensity.

{The mean \Ha\ intensity in the OFF NPS region with $\Tb\lesssim 35$ K is measured to be $\Iha \sim 1.22$ R, representing the mean Galactic emission \cite{haffner+2003} as indicated by the left side arrow marking the region
{in Fig. \ref{fig-tt}.}
The NPS shows up as the high-$\Tb$ extension at $35-60$ K {in the ON-NPS region marked by the right side arrow}.
So, we here define the NPS in this plot as the area with $\Tb (408 {\rm MHz})\gtrsim 35$ K, or $\Tb( 1420{\rm MHz})\gtrsim 1$ K for spectral index of $\beta=-2.7$.
We, then, measure the mean \Ha\ intensity in this ON-NPS region to be $\Iha \sim 1.3$ R.
Although the excess of ON-NPS $\Iha$ over OFF-NPS value is $\sim 0.1$ R, we here estimate the excess to be $\Delta \Iha\sim 0.2$R, considering the scatter and dispersion of the plotted values. 
This yields the \Ha-to-radio intensity ratio of
$\Q \lesssim 50$ at 1.4 GHz. }

The \Ha\ intensity yields the upper limit to the emission measure as $\EM\lesssim 0.5$ \EMunit for assumed electron temperature $\Te=10^4$ K. 
The values may be compared with those estimated for the SNR, Cygnus Loop,
where $\Tb\sim 1$ K and $\Iha\sim 20$ R, yielding $\Q\sim 6\times 10^3$, 
and $\EM\sim 50$ \EMunit for $10^4$ K as obtained from individual \Ha\ observations \cite{hester+1986}. 
We list the estimated values in table \ref{tabEM}. 
  
\section{Comparison with SNRs}

\subsection{Maps}

\begin{figure} 
\begin{center}     
\includegraphics[width=8cm]{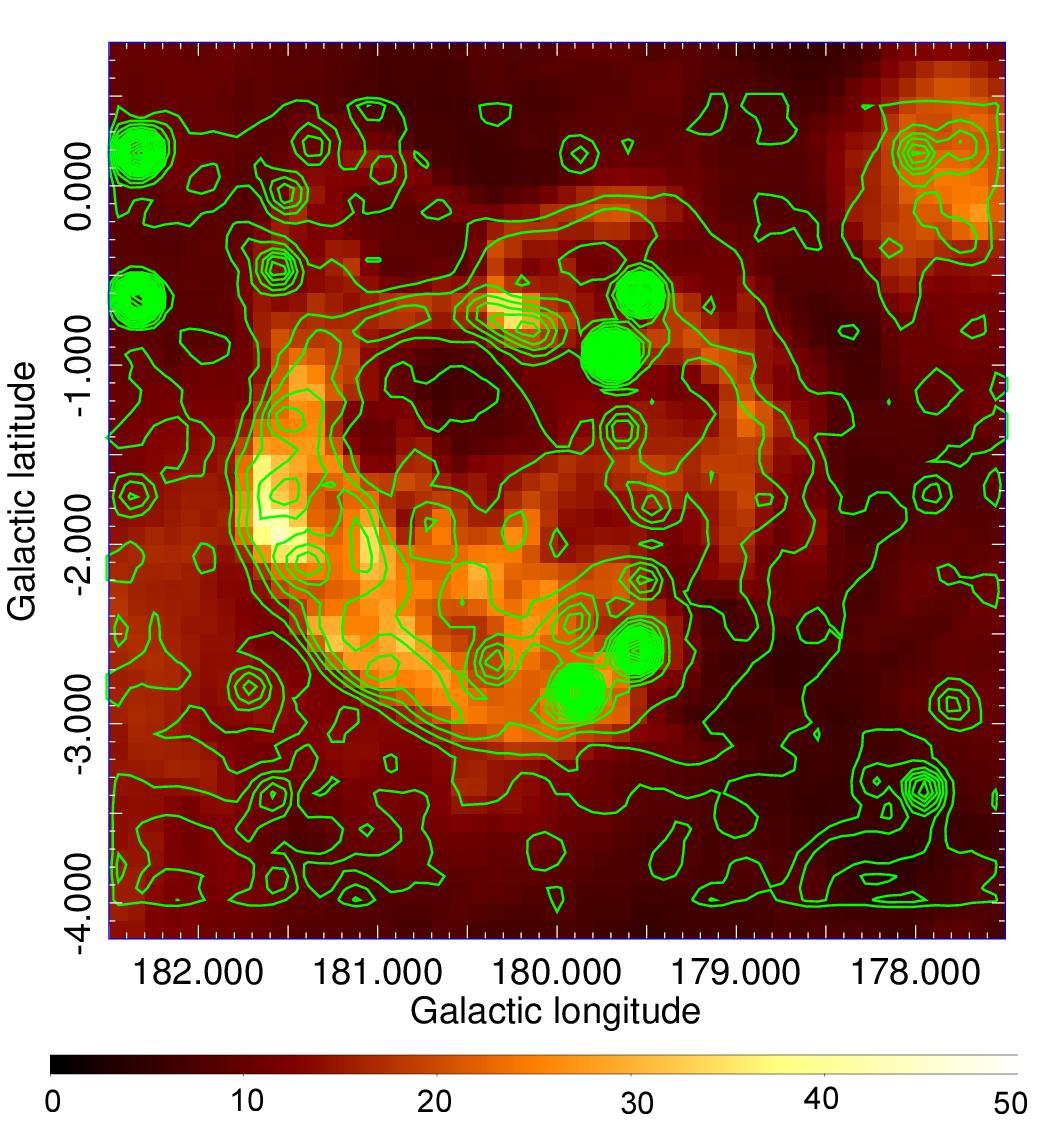}\\
\includegraphics[width=8cm]{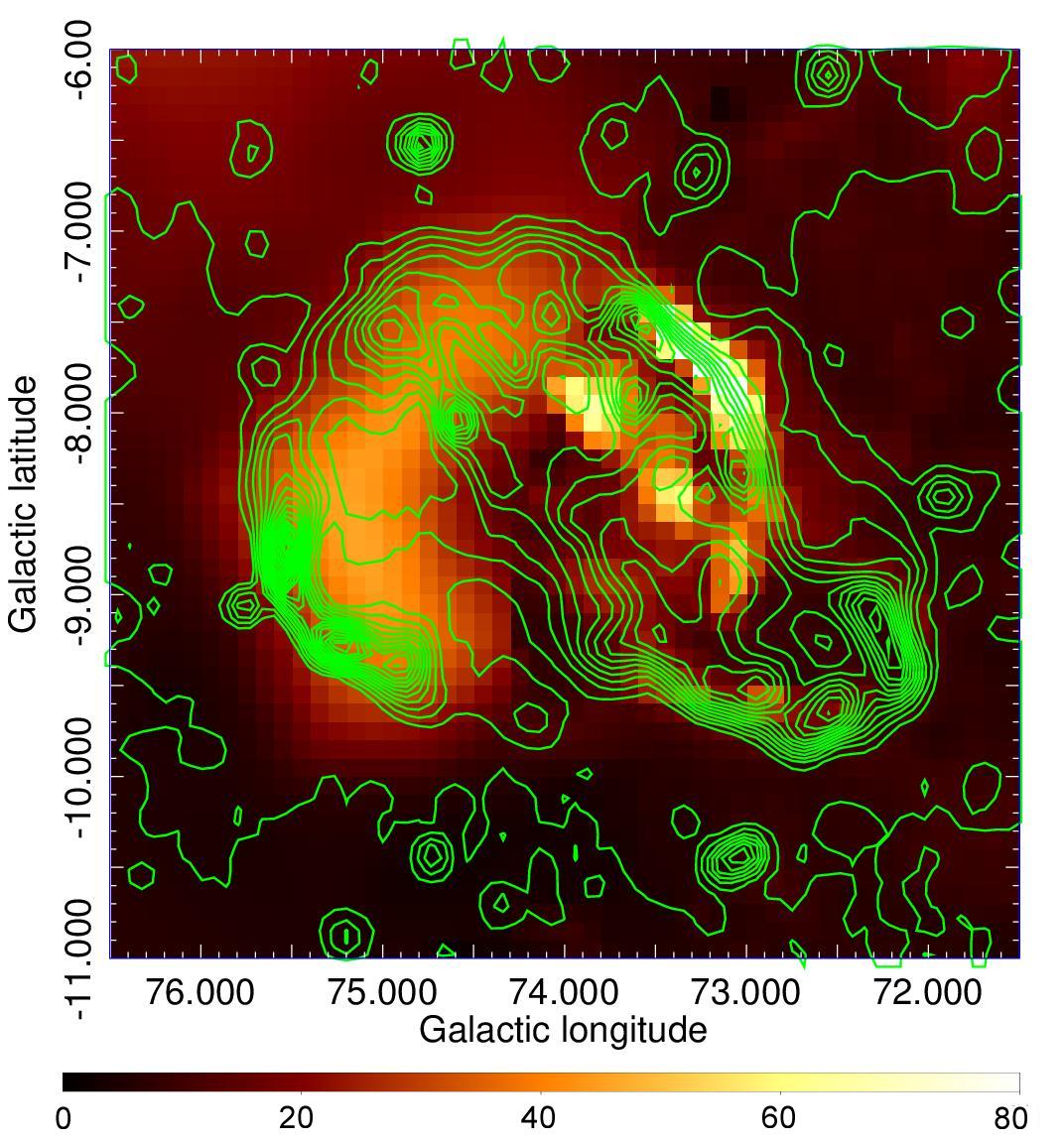}  
\end{center}   
\caption{  
1420 MHz maps of SNRs (top: S147, bottom: Cygnus Loop) from the Bonn-100 m Galactic plane survey
(Reich et al. 1990, 1997) by contours at an interval of 0.1 K overlaid on \Ha\ intensity maps from 0 to 50R for S147 and 0 to 80R for Cygnus Loop as indicated by the bars. 
Compare these maps with Figs. \ref{fig-allsky} and \ref{fig-nps-up}, where NPS is not visible in \ha\ despite the much lower intensity levels.} 
\label{fig-snr}  
\end{figure}

Fig. \ref{fig-snr} shows overlays of radio continuum brightness at 1420 MHz of the Cygnus Loop and S147  on \Ha\ intensity maps, where the radio maps were taken from the Bonn-100m Galactic plane survey \cite{reich+1990,reich+1997}.
The SNRs are visible both in radio and \Ha\ emissions.
Such \ha-radio association is {often} observed in Galactic radio SNRs \cite{uyaniker+2004,xiao+2008}, and 
seems also to happen in the spiral galaxy M31 \cite{braun+1993}. 
On the other hand, \Ha\ is hardly visible in the NPS despite the clear and sharp radio ridge as shown in Figs \ref{fig-allsky} and \ref{fig-nps-up}.
The NPS is, thus, extraordinarily fainter in \Ha\ compared with the usual SNRs.

\subsection{Spectra (SED)} 
\label{sec-spectra}

\begin{figure} 
\begin{center}  
\includegraphics[width=8cm]{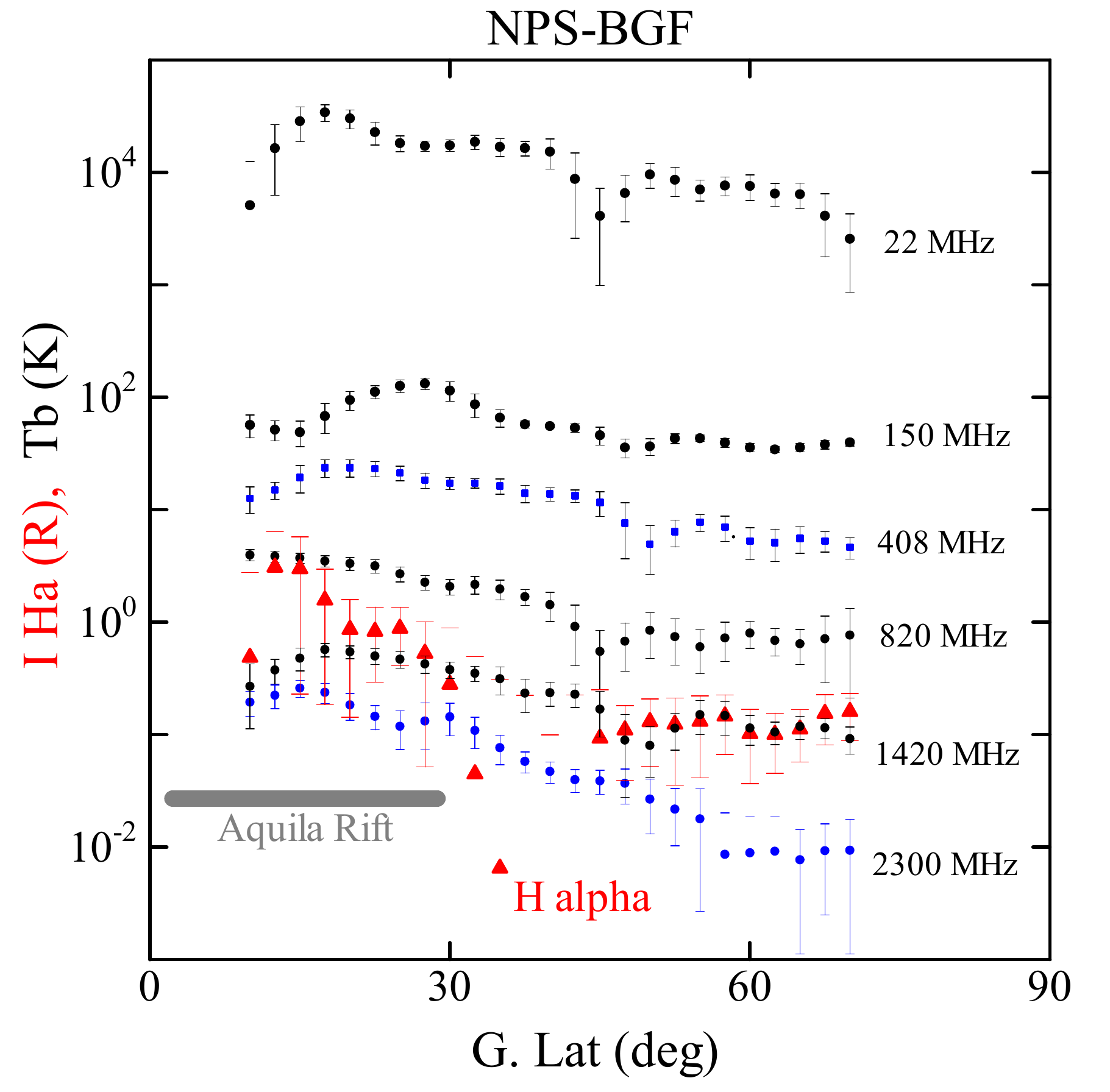}
\includegraphics[width=8cm]{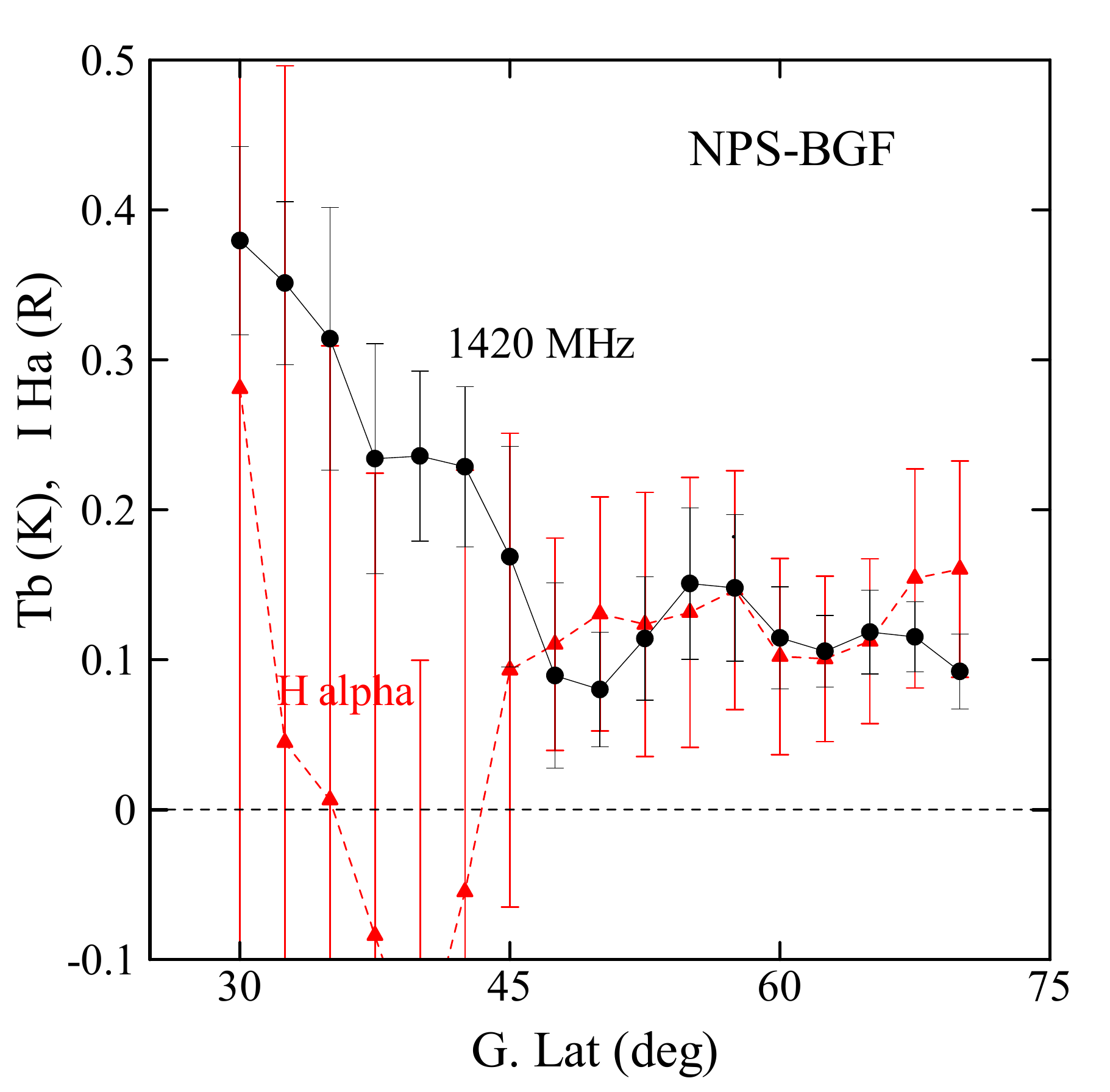}  
\end{center} 
\caption{[Top] Intensity profiles along the NPS ridge using the BGF maps at 22 to 2300 MHz and \ha\ maps taken from the surveys cited in section \ref{data}.
[Bottom] Same, but enlarged for 1420 MHz and \ha\ profiles.
}
\label{fig-profiles}  
\end{figure}

Fig. \ref{fig-profiles} shows variation of peak intensities at various frequencies along the NPS ridge as obtained using background-filtered (BGF) maps \cite{sofue+1979} of the radio and \ha\ sky surveys.
{The BGF subtracts background emission and creates a nearly zero-adjusted intensity distribution of sources with scale sizes greater than a smoothing beam width\footnote{{The 'background' is defined as a smoothed map after iterative clipping of peaky sources, so that it traces the valleys of the intensity distribution. 
}}}.
We used a box-shaped one-dimensional smoothing beam with full width of $10\deg$ in the longitude direction at each latitudinal grid.
The one-dimensional smoothing was so chosen that it avoids unnecessary smearing effect by the steep intensity gradient perpendicular to the Galactic disc.
Although the angular resolutions are different at different frequencies, the NPS is sufficiently resolved, and we used the peak intensities read on the thus obtained BGF maps for the spectral analysis in this section.

\begin{figure} 
\begin{center}     
\includegraphics[width=8cm]{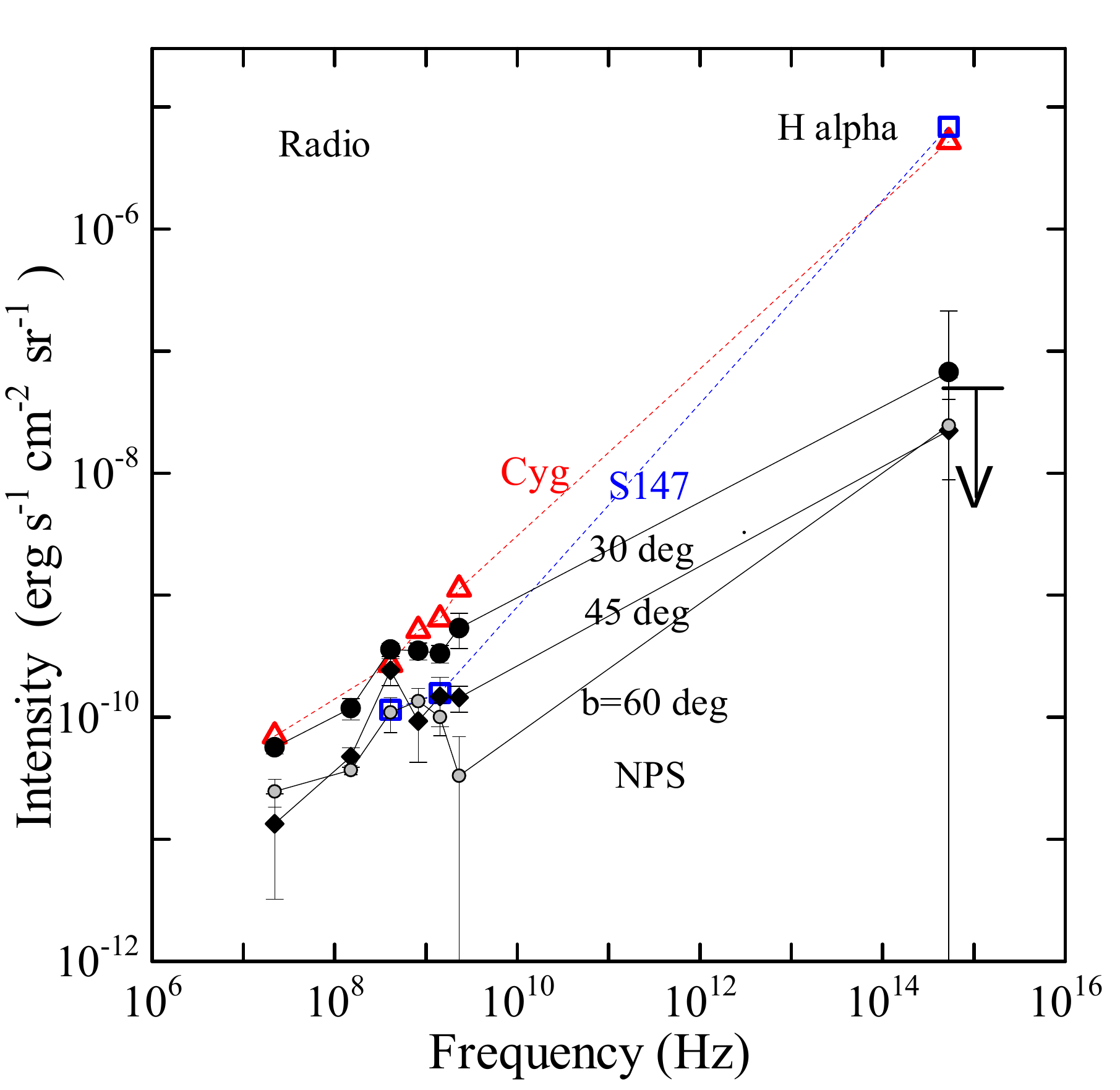} 
\end{center} 
\caption{Spectral energy distributions (SED) of the NPS at $b=30\deg$ (dots), $45\deg$ (diamonds), and $60\deg$ (grey circles) made from the peak-intensity profiles in Fig. \ref{fig-profiles}.
Intensities of the northern shell edge of the Cygnus Loop (red triangles) and eastern shell edge of S147 (blue squares) are also plotted. 
Interstellar extinction of \ha\ is corrected for according to their distances, while no correction is made for NPS here for its assumed distance of 140 pc in this diagram.
The arrow indicates the estimated upper limit of $\Iha$ from the $TT$ analysis.}
\label{fig-sed}  
\end{figure}

Fig. \ref{fig-sed} shows a spectral energy distribution (SED) of the NPS at $b=30\deg$, $45\deg$ and $60\deg$ as obtained from the intensity plot along the NPS ridge shown in Fig. \ref{fig-profiles}. 
We also plot SEDs of the northern shell edge of Cygnus Loop and eastern edge of S147.
The background emissions around SNRs are subtracted by measuring averaged brightness in 
{a small area without significant emission features}
about half a shell radius outside each shell edge at the same Galactic latitude.
The \ha\ extinction has been corrected by $A_{\rm H\alpha}=0.2$ and 0.6 mag., respectively, for Cygnus Loop and S147, as described in section \ref{sec-extinction}. 
However, the correction is not applied to the NPS, because we here compare the spectra, when they are assumed to be the same type (SNR-type) objects, and so the NPS is assumed to be located at a distance of $\sim 140$ pc with negligible extinction. 

{The radio spectra of NPS and SNRs are consistent with those from more accurate analyses \cite{iwashita+2023,xiao+2008,uyaniker+2004}, and show that the radio intensities of the NPS and SNRs are comparable.
However, a significant difference is found at \ha, where the NPS is fainter than SNRs by two orders of magnitude. 
The upper limit to the \ha\ intensity of the NPS as obtained from the $TT$ analysis is indicated by the arrow.}

\subsection{TT}

\begin{figure} 
\begin{center}     
\includegraphics[width=8cm]{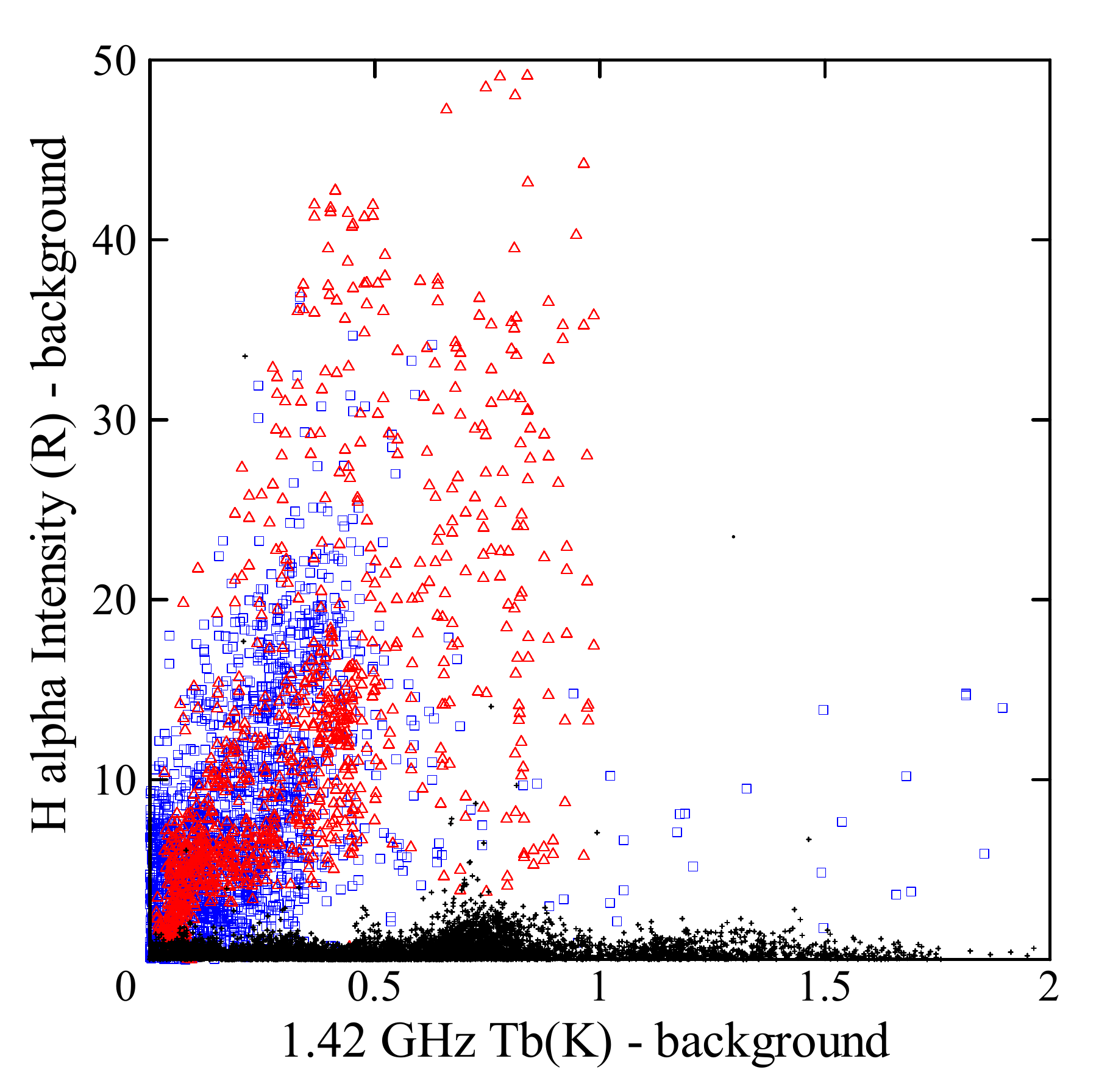}
\includegraphics[width=8cm]{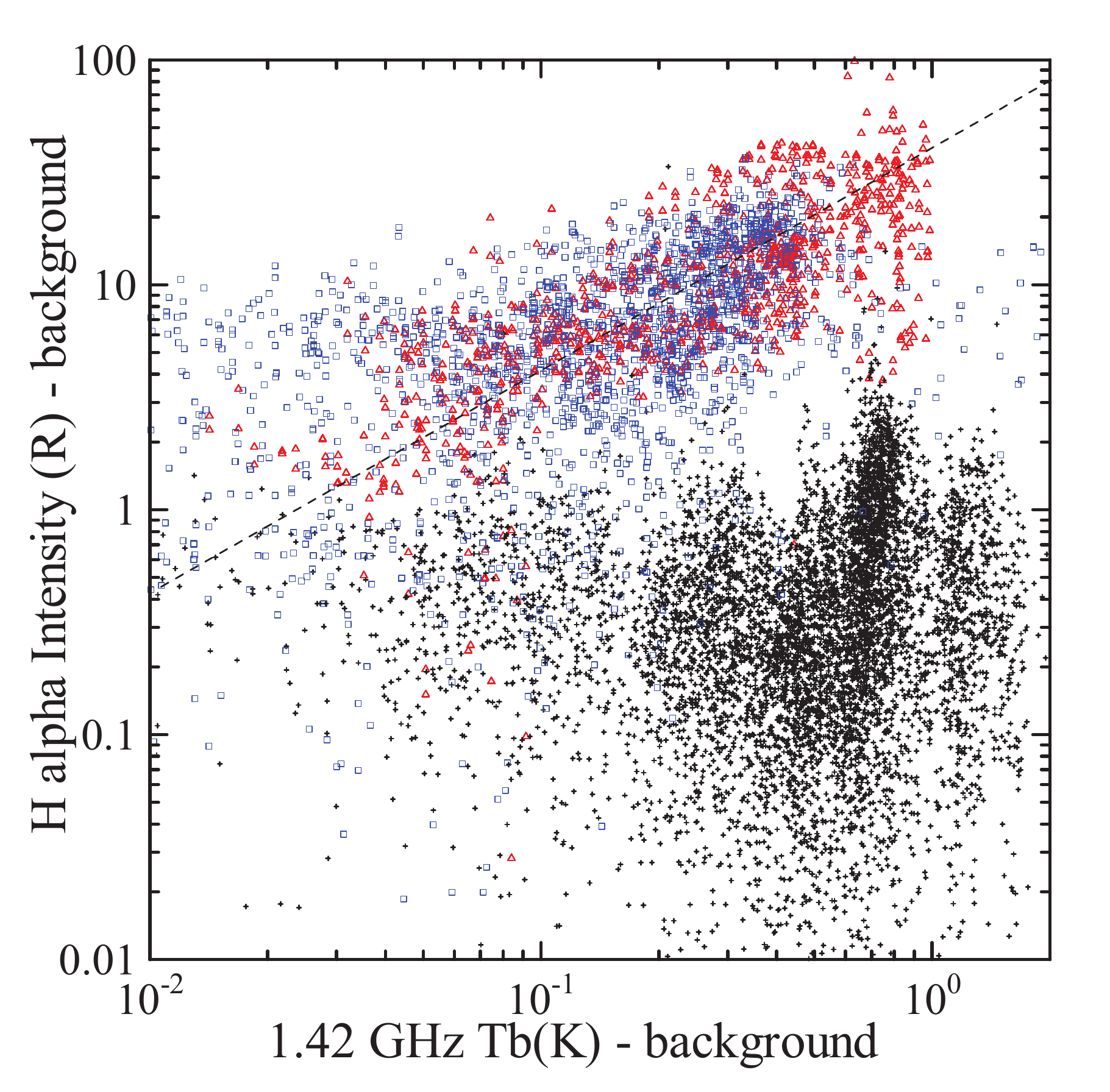}   
\end{center} 
\caption{Linear and logarithmic $TT$ plots of \Ha\ against radio intensities for the NPS (black dots) converted to 1.4 GHz compared with those at 1.4 GHz for Cygnus Loop (red triangles) and S147 (blue squares). 
Approximate base-line (background) intensities in each object's area are subtracted.
{The dashed line represents the \Ha-to-radio intensity ratio of $\Q\sim 1.1\times 10^4$.}}
\label{fig-tt-snr}  
\end{figure}

In Fig. \ref{fig-tt-snr} we display $TT$ plots in linear and logarithmic scaling for the SNRs shown in Fig. \ref{fig-snr} in comparison with that of the NPS in region T of Fig. \ref{fig-allsky}.
The lowest values of \Ha\ and radio intensities in each $TT$ plot representing the background emission are subtracted.
The plots for the SNRs indicate that the \Ha\ intensity is well correlated with the radio intensity by a relation $\Iha\propto \Tb$.
Using the gradient of the linear plot in the upper panel we obtain the \Ha-to-radio intensity ratio of $\Q \sim 1.1\times 10^4$ both for Cygnus Loop and S147. 
 On the other hand, the NPS's plot shows no clear correlation between \Ha\ and radio intensities.

\begin{table*}   
\caption{Intensities of the NPS and SNR Cygnus Loop}
\begin{tabular}{llllll}
\hline
\hline
Object &Quantity &   & EM (\EMunit)& $T_{\rm e}$(K) &Remark \\  
\hline
NPS & X ray (3/4 keV)&$I_{\rm X}\sim 2\times 10^{-4}$ cts/s/amin$^2$ &$\EM\sim 0.1$ & $\Te\sim 10^{6.5}$  (0.3 keV) &1, 2, 3\\
&\Ha\ for X-ray $\EM$ &$\Iha\sim 4\times 10^{-5}$ R& ibid ($ \sim 0.1$)  & for $10^{6.5}$ K  \\ 
&\Ha\ observed here &$\Iha\lesssim 0.2$ R &$\EM\lesssim 1.4\times 10^2$  & for $10^{6}$ K  \\ 
&\Ha\ observed here&$\Iha\lesssim 0.2$ R&$\EM\lesssim 0.4$  & for $10^4$ K  \\ 
& Radio 1.42 GHz&$\Tb\sim 1$ K\\
 &{{\bf \Ha/radio intensity ratio} }&{$\Q\lesssim 50$}\\
\hline
Cygnus Loop&Xray (3/4 keV)& $I_{\rm X}\sim 10^{-2}$ cts/s/amin$^2$&$\EM\sim 1$  & $\Te\sim 6\times 10^6$ (0.5 keV)&  4\\
&\Ha\ &$ \Iha\sim20-40$ R &$\EM\sim 50$& for $10^4$ K&5\\
& Radio 1.42 GHz & $\Tb\sim 1$ K\\
&{\bf \Ha/radio intensity ratio} & {\bf $\Q\sim 1.1\times 10^4$} &&& Fig. \ref{fig-tt-snr}.\\
\hline   
\end{tabular}\\
1. X-ray intensity was read from ROSAT all-sky map at 3/4 keV (R4 band) \cite{snowden+1997} 
(1 cts/s/arcmin$^2$ = 400 Jy sr$^{-1}$ $=9.6\times 10^{-4}$ erg cm$^{-2}$ s$^{-1}$ keV$^{-1}$);
2,3:\cite{kataoka+2013,yamamoto+2022}; 4. \cite{uchida+2008}; 5.\cite{hester+1986}. 

\label{tabEM}
\end{table*}  
\subsection{Interstellar extinction}
\label{sec-extinction}

The radio continuum emissions are absorption free in the circumstances discussed here regardless of the distance.
In the SNR hypothesis, which assumes a distance to NPS less than $\sim 140$ pc, the interstellar dust extinction of the \Ha\ emission is negligible in the entire Loop I even at low or zero latitudes, because it is located in front of the Aquila Rift at a distance of $\sim 0.4$ kpc \cite{sofue2015,sofue+2017}.
{This assumption is used in the figures where NPS and SNRs are compared.}

{The SNR Cygnus Loop is observed to have $A_v=0.25$ mag, or $A_{\rm H \alpha}\sim 0.2$ \cite{fesen+2018}.
S147 has foreground extinction of $A_v=0.7$ mag ($A_{\rm H \alpha}\sim A_r\sim 0.6$) \cite{fesen+1985}, and $A_v\sim 1.2$ mag or $A_r\sim 0.9$ including the internal extinction \cite{chen+2017}.
Therefore, the \Ha\ intensities observed toward Cygnus Loop and S147 in Fig. \ref{fig-snr} and \ref{fig-tt-snr} are under-estimated by about factors of 1.2 and 1.7, respectively. 
These factors for the SNRs are corrected in Fig. \ref{fig-sed} in order to compare with NPS as a local object with negligible extinction.}

In the GC explosion hypothesis of NPS, which assumes a distance of $\sim 8$ kpc, the optical (visual) extinction can be estimated by the general law relating it to HI column density, $N_{\rm HI}$, by
\begin{equation}
A_v=N_{\rm HI}/1.79\times 10^{21}{\rm HI\ cm^{-2}}
\end{equation} 
\cite{predehl+1995}.
Measuring the HI column density from the all-sky integrated HI intensity map \cite{kalberla+2007}, we obtain 
$N_{\rm HI}\sim 1.1\times 10^{21}$ H cm$^{-2}$ at $b\sim 15\deg$,
$\sim 0.7\times 10^{21}$ at $30\deg$, and
$\sim 0.15\times 10^{21}$ at $\sim 60\deg$.
Then, assuming \cite{gordon+2003}
\be
A_{\rm H \alpha}\simeq A_{\rm r}=0.8 A_{v},
\ee
we obtain 
$A_{\rm H \alpha}\sim 0.49$, 
0.31 and $\sim 0.07$ mag. at 
$b\sim 15\deg$, $30\deg$, 
and $60\deg$, respectively.
Or, the \Ha\ intensities are under-estimated by a factor of 1.6, 1.3 and 1.07, respectively, at these latitudes.  
The region closer to the low galactic latitudes of NPS suffers from heavier extinction by the dust lane of the Aquila Rift, where we cannot give a conclusive discussion.

\section{Summary and Discussion} 

\subsection{Summary}
Analysis has shown that the {\Ha}-to-1.4 GHz intensity ratio for NPS ($\Q\lesssim 50$) is more than two orders of magnitude smaller than that for typical shell-type SNRs, Cygnus Loop and S147 ($\Q\sim 1.1\times 10^4$). 
No evidence of \ha\ association was found along NPS, even towards the brightest and sharpest ridge at $b\sim 8-20\deg$.
The low \ha\ intensity favours the GC explosion model, which postulate the distance of $\sim 7$ kpc, over the local supernova explosion model.   

Below we discuss the implication of the results for the two models about the origin of NPS. 
In either model of local or GC origin, it should be noted that NPS is much larger in size than the other known SNRs or bubbles.
This may lead to various differences in environments and physical conditions where the NPS is situated. 
The $\Q$ value would be useful to distinguish such differences from each other regardless of the size and distance.
 
\subsection{On the local bubble model} 

{It is difficult to interpret the H$\alpha$-dark NPS as an ordinary supernova remnant such as the Cygnus Loop or S147 that exploded in the dense ($\rho\sim 1$ H cm$^{-3}$) and cold ($ T\lesssim 10^4$ K) Galactic disc, where the hydrogen recombination is high and the shock compressed shells efficiently emit \Ha\ line at $\sim 20-40$ R.  
If NPS is a remnant of multiple supernovae exploded in a local OB association, much stronger shock waves would cause brighter \Ha. 
If it is a similar object to the Orion-Eridanus super bubble \cite{pon+2016} as seen in Fig. \ref{fig-allsky} around $(l,b)\sim (200\deg,-20\deg)$, it should be as bright as $\sim 10-20$ R. 
If it is associated with high-latitude HI spurs at local velocities \cite{heiles+1980}, then the shocked area in touch with the NPS should emit \Ha\ by the same mechanism as above. 
Thus, the local origin models have difficulty explaining} the \ha\ faintness of NPS.

\subsection{{On the GC explosion model}}

{The \Ha\ faintness can be naturally explained by the GC explosion model.
The model postulates a shock wave propagating in the Galactic halo at $T\sim 10^{6.3}$ K.
The temperature in NPS is observed to be much higher at $\sim 10^{6.5}$ K, while the emission measure is rather small at $\EM\sim 0.1$ \EMunit \cite{kataoka+2013,kataoka+2018,yamamoto+2022}.
In such a hot plasma the hydrogen recombination rate is $10^{-2}$ times that in the ISM at $T\sim 8000$ K, $\alpha(10^{6.5} {\rm K}) \sim 10^{-2} \alpha(8000{\rm K})$  \cite{hummer1994}.
Knowing that 1 R at this temperature corresponds to 2.25 \EMunit \cite{haffner+2003}, it leads to 
$\Iha\sim 10^{-2} EM/2.25 \sim 4\times 10^{-3}$ R, 
which is consistently below the observed upper limit of $\IhaNPS \lesssim 0.2$ R.
Thus, the GC bubble model seems plausible to explain the observed \ha\ faintness of the NPS. }

{The \ha\ property of NPS gives further constraint on the GC origin model.
A galactic-scale wind such as observed in starburst galaxy NGC 3079 \cite{cecil+2002} or M82 \cite{lehnert+1999} presumes a large \ha\ shell in the halo.
However, such an \ha\ shell is not observed in NPS.
So, NPS may be a more spherical bubble directly exposed to the halo’s hot plasma.
Even so, the root region might emit \Ha\ associated with the 1-kpc conical wind of HI \cite{sofue2022hi,sofue+2021} and X-ray \cite{bland+2003}, but it is hidden behind the Aquila Rift. 
In this case the NPS will be a giant bubble similar to that observed in NGC 253, where 1-kpc scale \Ha\ wind blows near the nucleus \cite{west+2011} and a giant X-ray and radio bubbles are expanding into the halo \cite{sofue+2001}.}

\section*{Acknowledgements} 
The data analysis was carried out on the computer system at the Astronomy Data Center of the National Astron. Obs. of Japan.  
The authors are indebted to the groups of the Bonn-Stockert radio (Dr. P. and W. Reich) and WHAM (Dr. L. M. Haffner) surveys for the archival data base.
A part of this research was supported by Japan Science and Technology Agency (JST) ERATO Grant Number JPMJER2102 and JSPS Kakenhi Grant Number 20K20923, Japan.
 
\section*{Data availability}  
The radio and \Ha\ fits data were downloaded from the URL:\\
https://lambda.gsfc.nasa.gov/product/foreground/ fg\_diffuse.html, 
and  \\
http://www3.mpifr-bonn.mpg.de/survey.html.
  
\section*{Conflict of interest}
The authors declare that there is no conflict of interest.


\end{document}